\documentclass[preprint,showpacs,preprintnumbers,amsmath,amssymb,nofootinbib]{revtex4}
\usepackage{graphicx}
\usepackage{dcolumn}
\usepackage{bm}

\begin{document}
\title{\bf Classic field theories of gravitation embedded in ten dimensions}
\author{Frank B. Estabrook}
\email{frank.b.estabrook@jpl.nasa.gov}
\affiliation{Jet Propulsion Laboratory, California Institute of Technology, Pasadena, CA 91109}

\date{\today}

\begin{abstract} Two classic field theories of metric gravitation are given as constant-coefficient Exterior Differential Systems (EDS) on the flat orthonormal frame bundle over ten dimensional space.  They are derivable by variation of Cartan 4-forms, and shown to be well-posed by calculation of their Cartan characteristic integers. Their solutions are embedded Riemannian 4-spaces.  
The first theory is generated by torsion 2-forms and Ricci-flat 3-forms and is a constant-coefficient EDS for vacuum tetrad gravity; its Cartan character table is the same as found for an EDS recently given in terms of tetrad frame and connection variables \cite{estabrook1} \cite{estabrook2}.  The second constant-coefficient EDS is generated solely by 2-forms, and has a Cartan form of quadratic Yang-Mills type. Its solutions lie in torsion free 6-spaces and are fibered over 3-spaces.  We conjecture that such solutions may be classically related to 10-dimensional quantum field theoretic constructions of cosmological vacua \cite{gherghetta}.

\end{abstract}
\maketitle
\baselineskip= 14pt
\section{Introduction} A metric theory of gravitation describes the Riemannian geometry of four (3+1) dimensional spacetime.  Free (test) particles, subject only to gravitation, move on geodesics, and inertia and the equivalence principle are thus made intrinsic.  Its connection variables are the gravitational field, and their partial differential equations can follow from a variational principle. Cartan's method of the movable frame can be used to formulate a metric theory as a classic field theory, a well posed set of first order partial differential equations derivable from a functional Lagrangian, or equivalently as an Exterior Differential System (EDS) together with a Cartan form. One instance of such a program \cite{estabrook1} was the formulation of vacuum general relativity using tetrad vector components and Ricci rotation coefficients as 40 dependent fields, together with 4 independent spacetime coordinates, so the EDS and Cartan form were set in terms of 40 variables (scalar fields that entered quadratically) and 44 basis 1-forms. Ten conservation laws were found (technically, arising from isovectors forming 10 dimensional fibrations in solutions), leaving in principle 34 bases for calculation of the Cartan integer characters showing the well-posed nature of the EDS \cite{estabrook2}. In Section II we will present a new EDS for vacuum tetrad gravity using a larger number of basis forms (55), but set as a so-called constant-coefficient or c-c ideal \cite{ccideals} in which outer products of bases occur with purely numerical coefficients (no coordinate functions).  It is abstractly equivalent to the former EDS, as now there are 21 Cauchy characteristics (fibers of the EDS) and so again just 34 remaining bases, and the calculated Cartan characteristic integers come out the same. A c-c ideal is considerably more elegant and algebraically efficient for field theoretic derivations.

The most fundamental setting of an EDS for four dimensional movable frame geometry is the \textit{flat} orthonormal frame bundle over 10 dimensions.  Ten dimensions is the maximum needed for embedding of an arbitrarily curved 4-space in a flat embedding space.  The known exact solutions of the Einstein equations have in fact been classified as of classes 1-6 depending on whether they can be embedded in \textit{flat} spaces (taken with compatible signature) of dimension 5-10 respectively \cite{Stephani}. Movable frames in flat ten dimensional spaces are cross sections of an othonormal frame bundle with the structure of a Lie group of isometries of flat ten dimensional space.  The fibers are 45 dimensional, each the rotation group $O(10)$ (or with different signature $O(9,1)$, etc.) The canonical 1-forms are grouped into ten bases denoted ${\theta _\mu, \mu = 1,..10}$ (in a cross section these become an orthonormal frame, a decad), and 45 connection bases ${ \omega_{\mu \nu} }$ antisymmetric in ${ \mu ,\nu }$. The structure equations are 
\begin{align}
{d\theta }_{\mu }+{\omega }_{\mu \nu} \wedge {\theta }_{\nu } &= 0 \\
{d\omega }_{\mu \nu}+{\omega }_{\mu \sigma}\wedge {\omega
}_{\sigma \nu} &= 0.
\end{align}
The zero on the right hand side of Eq.(1) says the \textit{torsion} of the basis ${\theta_\mu}$ vanishes, a requisite for Riemannian geometry,  and the zero on the right hand side of Eq.(2) says the 45 Riemann 2-forms, or \textit{curvature} tensor, vanish. As a result, Eq.(1) and Eq.(2) are Cartan-Maurer equations of  55 dimensional groups denoted $ISO(10)$ or $ISO(9,1)$, etc. (The  decad indices ${ \mu , \nu }$ can be raised and lowered with a constant symmetric 10 x 10 matrix ${\eta ^{\mu \nu }}$ to allow for various signatures; for convenience we place all indices downstairs, and remember that the ${\eta ^{\mu \nu }}$ must be used when summing repeated indices.)  Coordinate expressions for the 55 basis 1-forms (a representation of the Lie group) can in principle be found, similarly to the use of 3 translations and 3 Euler angles in the familiar bundle of frames over Euclidean 3-space; then the structure equations of the bases become trivial, but the c-c structure of the ideal is lost.  

It is the use of the anholonomic bases ${\theta _\mu,   \omega_{\mu \nu} }$, together with the structure Eqs.(1) and (2), that allows us in Section II to set an embedding EDS for vacuum gravitation theory as as a c-c ideal. Its Cartan form is a 4-form for the Hilbert action.  In Section III we present a new embedding c-c EDS for a field theory whose Cartan form is of Yang-Mills type, and whose solutions may be classical analogues of brane-theoretic cosmological vacua.  From an exploration of possible EDS' for four dimensional field theories set using the embedding approach, these two well posed theories appear to be unique.  
\section{Tetrad Gravity}
 We will write the two EDS' for embedding using partitions $(n,
m),  n + m = 10$, of the basis forms of $ISO(10)$ into classes labeled
respectively by the first $n$ indices $i,  j$,  etc. $= 1,  2,  ... n$ and the
$m$ remaining indices $A,  B$,  etc. $= n +1,  n +2,  ... 10$.  So the basis
forms are rewritten as ${{\theta }_i},  {{\theta }_A},  {{\omega }_{{ij}}} = - {{\omega
}_{{ji}}},  {{\omega }_{{AB}}} = - {{\omega
}_{{BA}}},  {{\omega }_{{iA}}} =-{{\omega }_{{Ai}}}$.  Summation
conventions on repeated indices will be used separately on each partition.  
The structure equations (1), (2) become 
\begin{align}
{d\theta }_{i}+{\omega }_{ij} \wedge {\theta }_{j} &= -{\omega }_{iA}
\wedge {\theta }_{A}  \\
{d\theta }_{A}+{\omega }_{AB} \wedge {\theta }_{B} &= - {\omega }_{Ai}
\wedge {\theta }_{i} \\
{d\omega }_{ij}+{\omega }_{ik} \wedge {\theta }_{kj} &= -{\omega }_{iA}
\wedge
{\omega }_{Aj} \\
{d\omega }_{AB} +{\omega }_{AC} \wedge {\omega }_{CB} &= {\omega }_{iA}
\wedge
{\omega }_{iB} \\
{d\omega }_{iA} +{\omega }_{ij} \wedge {\omega }_{jA} + {\omega }_{AB}
\wedge
{\omega }_{iB} &= 0.
\end{align}
We have moved to the right of the zeros terms that in the following will be interpreted as torsions and curvatures of submanifolds; as for surfaces in Euclidean 3-space, these are local properties induced by simple embedding.  We do not treat global or intrinsic topological complications, which is why 10 embedding dimensions suffice.

 The  EDS of vacuum tetrad gravity is set using the partition (4,6), i.e. i,j, etc. = 1,..4,   A,B, etc. = 5,..10, and the setting of the Hilbert action is by means of the following Cartan 4-form 
\begin{equation}
\Lambda = \omega_{iA}\wedge \omega_{jA}\wedge \theta_k\wedge \theta_l
\epsilon^{ijkl},
\end{equation}
From the Gauss structure equation,  Eq. (5), we recognize that $2{R_{{ij}}} = {{\omega }_{{iA}}}\wedge \)\({{\omega }_{{jA}}}$ will become the induced curvature when pulled back to a 4-dimensional submanifold, and $\Lambda$ will be the Ricci scalar.  If the induced torsion, from the structure equation Eq.(3), $T_i = -{\omega }_{iA}
\wedge {\theta }_{A}$, also vanishes we have a Riemannian submanifold.

 The exterior derivative of the $4$-form field $\Lambda$ on $ISO(10)$,  using Eq. (3) and (7),  is quickly calculated to be the 5-form
\begin{equation}
d \Lambda = T_i \wedge R_{jk} \wedge \theta_l \epsilon^{ijkl}.
\end{equation}
We have in \cite{estabrook1} discussed how the variation of a Cartan form $\Lambda$ leads to an EDS by quadratic factoring of the multisymplectic form $d \Lambda$.  Factoring Eq.(9) gives an EDS that is formally the same as found in \cite{estabrook1}:
\begin{equation} \{ T_i, R_{ij} \wedge \theta_k \epsilon^{ijkl}, R_{ij} \wedge \theta_j \}
\end{equation}
$ R_{ij} \wedge \theta_k \epsilon^{ijkl}$ is the Ricci 3-form (equivalent to the Ricci tensor), and since also $T_i$ will vanish on solutions, the latter will be (frame bundles over) vacuum Riemannian 4-spaces.  The second set of 3-forms must be included in the EDS for closure (they are $dT_i$).

The first new result presented in the present paper is the Cartan analysis \cite{Ivey} of the EDS generated by Eq.(10).  In this research we use the AVF form manipulation program of H. D. Wahlquist , together with his Monte-Carlo program for calculating Cartan's integer characters \cite{Wahlquist}.  With it he found for the tetrad gravity EDS Eq.(10) the character table $55\{0,4,12,14\}4+21$ (with the conventions of \cite{estabrook1}).  This satisfies Cartan's criteria \cite{Ivey}: this field theory is well posed, has Cauchy-Kovalevskaya properties of local existence and uniqueness of solutions, so is a causal physical theory (if 3+1 signature is adopted); the $\theta_i$ are independent (involutory) and become an orthonormal tetrad basis in a solution. $\omega_{ij}$ and $\omega_{AB}$ do not appear in the EDS,  so there are 21 Cauchy characteristics corresponding to 6 free rotations of the tetrad frame $\theta_i$ and 15 free rotations of the coframe forms $\theta_A$.

It should be remarked that this dynamic field theory has more algebraic elegance, and probably physical  interest, than the more usual mathematical setting of \textit{isometric} embedding \cite{Griffiths}.  In problems where Riemannian manifolds are embedded in higher dimensional \textit{nonflat} Riemannian spaces, the coframes $\theta_A$ are also required to vanish on the submanifold.   In our formalism the $\theta_A$ remain, but pulled back as fields on the solutions having independent interest.

\section{Double Torsion Free Theory}The tetrad vacuum gravitation EDS of Section II had four 2-forms and eight 3-forms in its generating ideal. An EDS generated only by an ideal of 2-forms is dual to an incomplete Lie operator algebra of finite growth; it can be of Kac-Moody type, and have mathematically elegant integrability and nonlinear superposition properties \cite{ccideals}). Motivated by this, we have searched whether just the torsion 2-forms induced in \textit{both} the local partitions, Eq.(3) and Eq.(4), can together be taken as generators of an interesting EDS:
\begin{equation} 
\{\omega_{iA} \wedge \theta_A , \omega_{iA} \wedge \theta_i \}
\end{equation}
It can easily be checked that it is closed. We discovered by calculation of the characteristic integers of the ideal Eq.(11) for many cases that \textit{iff} we use the partition (3,7) the EDS is well posed with genus 4.  That is, we must take in Eqs.(3)-(7) the range i = 1, 2, 3 and A = 4,..10 (or conversely).  The Cartan character table of the EDS generated by Eq.(11) then is well posed: 55\{0,10,9,8\}4+24. This is the second new result of this paper. $\omega_{ij}$ and $\omega_{AB}$ do not appear in the EDS, so there are Cauchy characteristic fibers of dimension 24 (3 from $O(3)$ plus 21 from $O(7))$.  In a solution, the three $\theta_i$ and one of the $\theta_A$, or any four of the $\theta_A$, are in involution, but we refrain from calling any of these sets an orthonormal basis there.  The solutions are $O(3)\otimes O(7)$ bundles over 4-space deriving their metric geometry as subbundles of $ISO(10)$.  The induced covariant metric tensor is the pullback of $\theta_i \theta_i + \theta_A \theta_A$ to a cross section. 

A Cartan 4-form for this theory (i.e., a variational principle) can also readily be found. It is quadratic in the torsion 2-forms $T_i$ in Yang-Mills fashion:
\begin{equation}
\Lambda = T_i \wedge T_i.
\end{equation}
Equivalently, up to an exact (boundary) term, we could have used the second torsion, say  $T_A = -\omega_{ iA}\wedge {{\theta }_i}$ from Eq.(4), and set
\begin{equation}
\Lambda = T_A \wedge T_A.
\end{equation}
The multisymplectic 5-form from either is
\begin{equation}
d \Lambda = T_i \wedge \omega_{iA} \wedge T_A
\end{equation}
which factors to give the EDS Eq.(11).

Some details of this result can be further interpreted geometrically.  The solution manifolds are $O(3)\otimes O(7)$ bundles over 4 dimensions; they are subbundle maps into the $O(10)$ frame bundle over the flat 10-dimensional embedding space. The closed structure equations of the three forms $\theta_i$ show them to define a vector field along which (together with the fiber $O(3) \otimes O(7)$) there is a map to a quotient space, $4 \rightarrow 3$. This is sometimes called a Riemannian submersion or rigid motion congruence. $R_{ij}$ is the resulting 3-dimensional curvature there, and $R_{ij}\wedge \theta_k \epsilon_{ijk}$ the scalar curvature.  This 3-dimensional metric geometry does not appear to be algebraically special, but we have found, using Wahlquist's Monte Carlo solutions of the EDS, that the 4-form $R_{ij}\wedge R_{ij}$ does vanish on all solutions.  Since this expression is the exterior derivative of the so-called Chern-Simons 3-form $\omega_{ij} \wedge \omega_{jk} \wedge \omega_{ki} + 6 \omega_{ij} \wedge R_{ij}$, the latter is a conservation law of the (4-dimensional) solution.  

There are also interesting inclusion maps.  The double torsion free EDS Eq.(11) is contained in the c-c EDS for isometrically embedded flat 3-spaces $\{ \omega_{iA} , \theta_A \}$, which has the well posed character table $55 \{0,28,0,0 \}3+24$.  Also the EDS for torsion-free co-frames,  $\{T_A, dT_A\}$, that is contained in the double torsion free EDS Eq.(11), is well posed with character table $55 \{0,7,14,4,0,0\}6+24$. Its solutions are $O(3)\otimes O(7)$ bundles over embedded 6-spaces.  So these nested EDS' result in a map of metric spaces $3\rightarrow 4 \rightarrow 6\rightarrow 10$. 
   
 Without belaboring the differential geometry further, we claim the importance of this must lie in the recognition that it gives well posed classical dynamics; we emphasize that this double torsion free EDS fulfills all Cartan's criteria for a field theory in four dimensions to be well posed, exactly as does the conventional vacuum tetrad relativity of Section II. Specifically, these are the conditions for the theory to have Cauchy-Kovalevskaya properties of local existence and uniqueness. The theory is surely nonlinear, as is vacuum general relativity, but since it is generated only by 2-forms, superpositions and transformations among solutions may be more tractable. We are aware of possible parallels with recently suggested 10 dimensional quantum field theories of cosmological vacua (we cite just two examples from a very large literature \cite{gherghetta}), in which interaction of variously dimensioned embedded branes in early epochs evolves physical gravity. Work is continuing on this, together with understanding double torsion free embedding theories set in different dimensions, such as 3 dimensional double torsion free embedding into flat 6, 5 into 15, and so on.

As a final remark, we have performed extensive Cartan character calculations on many other possible and plausible closed EDS', all set, as in the foregoing, with the induced torsion and curvature forms that arise from various partitionings, Equations (3)-(7).  The above two EDS', belonging to partitions (4,6) and (3,7) respectively, were the only ones found that result in four embedded dimensions and that satisfy Cartan's criteria for being well posed.

\subsection{Acknowledgements}

This research was performed while the author held a visiting appointment at the Jet Propulsion Laboratory,  California Institute of Technology

\end{document}